\documentclass[aps,prb,twocolumn,showpacs,citeautoscript,reprint,longbibliography,superscriptaddress]{revtex4-1}
\usepackage{graphicx}
\usepackage{bm,amsmath,amssymb,wasysym,amsfonts,dcolumn}
\usepackage[colorlinks=true, urlcolor=blue, linkcolor=blue, citecolor=blue, pdfborder={0 0 0}]{hyperref}
\usepackage[usenames,dvipsnames]{xcolor}
\usepackage{mathrsfs}

\newcommand{\be}{\begin{equation}}
\newcommand{\ee}{\end{equation}}
\newcommand{\bea}{\begin{eqnarray}}
\newcommand{\eea}{\end{eqnarray}}

\begin{document}

\title{Ultrafast magnetization enhancement and spin current injection in magnetic multilayers by exciting the nonmagnetic metal}
\author{Wen-Tian Lu}
\affiliation{Key Laboratory of Magnetic Molecules and Magnetic Information Materials of Ministry of Education $\&$\\ School of Chemistry and Materials Science of Shanxi Normal University, TaiYuan 030032, China}
\author{Zhe Yuan}
\email[Corresponding author: ]{yuanz@fudan.edu.cn}
\affiliation{Institute for Nanoelectronic Devices and Quantum Computing, Fudan University, Shanghai 200433, China}
\author{Xiaohong Xu}
\email[Corresponding author: ]{xuxh@sxnu.edu.cn}
\affiliation{Key Laboratory of Magnetic Molecules and Magnetic Information Materials of Ministry of Education $\&$\\ School of Chemistry and Materials Science of Shanxi Normal University, TaiYuan 030032, China}

\date{\today}

\begin{abstract}
A systematic investigation of spin injection behavior in Au/FM (FM = Fe and Ni) multilayers is performed using the superdiffusive spin transport theory. By exciting the nonmagnetic layer, the laser-induced hot electrons may transfer spin angular momentum into the adjacent ferromagnetic~(FM) metals resulting in ultrafast demagnetization or enhancement. We find that these experimental phenomena sensitively depend on the particular interface reflectivity of hot electrons and may reconcile the different observations in experiment. Stimulated by the ultrafast spin currents carried by the hot electrons, we propose the multilayer structures to generate highly spin polarized currents for development of future ultrafast spintronics devices. The spin polarization of the electric currents carried by the hot electrons can be significantly enhanced by the joint effects of bulk and interfacial spin filtering. Meanwhile the intensity of the generated spin current can be optimized by varying the number of repeated stacking units and the thickness of each metallic layer.
\end{abstract}

\maketitle

\section{Introduction}\label{chap1}

Microelectronics in the post Moore era is facing significant challenges and opportunities. One of the critical challenges in magnetic storage technique is known as ``the memory wall problem'' \cite{jhang2021challenges, zou2021breaking, fouda2022in}, which arises from the much lower speed of reading and writing digital information than the processing in the computing units. Typical devices for data storage in spintronics exploit different magnetic states to represent ``0'' and ``1'' \cite{wolf2001spintronics, chappert2007the, grollier2020neuromorphic}, whereas the transition between different states usually require at least nanoseconds \cite{hayakawa2021nanosecond}. With the reported phenomenon of femtosecond laser induced ultrafast demagnetization in nickel nanofilms \cite{beaurepaire1996ultrafast} and ultrafast magnetic switching \cite{kirilyuk2010ultrafast}, ultrafast spin dynamics on the subpicosecond time scale has become a promising solution for the development of the next generation magnetic storage devices\cite{lu2022progress}.

Multiple microscopic mechanisms based upon local spin-flip scattering have been proposed to explain the efficient dissipation of spin angular momentum in ultrafast demagnetization \cite{zhang2000laser, koopmans2005unifying, carpene2008dynamics, krauss2009ultrafast, bigot2009coherent, qaiumzadeh2013manipulation, tveten2015electron, freimuth2021laser}. Koopmans {\it et al.} introduced the Elliot-Yafet type electron-phonon spin-flip scattering into the phenomenological three-temperature model and established a microscopic three-temperature model \cite{koopmans2010explaining}. Later, experimentalists have shown that spin angular momentum also dissipates via nonlocal spin current transport in addition to the local spin flip scattering. Malinowski {\it et al.} first discovered that interlayer transport of spin-polarized hot electrons is closely related to ultrafast demagnetization induced by laser pulses in magnetic multilayers \cite{malinowski2008control}. Subsequently, Battiato {\it et al.} theoretically proposed the superdiffusive spin transport model, in which the spin current carried by the laser-excited hot electrons plays a major role in the ultrafast spin dynamics \cite{battiato2010superdiffusive}. Soon after that, the nonlocal spin transport were found to be remarkably important in many ultrafast phenomena of magnetism and spintronics, such as the ultrafast demagnetization induced by nonmagnetic (NM) metal excitation \cite{eschenlohr2013ultrafast}, and spintronic terahertz emission \cite{kampfrath2013terahertz, liu2021strain}. In a Ni/Ru/Fe trilayer with parallel magnetization in both FM metals, Rudolf {\it et al.} found that laser-induced demagnetization in Ni instantaneously enhanced the magnetization of the Fe layer, indicating interlayer spin transfer carried by hot electrons. It is expected that large ultrafast spin transfer with a magnetic moment can be applied in the motion of domain walls, ultrafast magnetic switching, and terahertz spintronics \cite{rudolf2012ultrafast}. In addition to conducting metals, hot electrons also generate tunneling spin currents in insulating materials. He {\it et al.} found that the ultrafast demagnetization process in a CoFeB/MgO/CoFeB magnetic tunnel junction can be controlled by hot electron tunneling \cite{he2013ultrafast, ji2023ultrafast}. Thus far, experimental observations have revealed that laser excitation of an FM metal can produce interlayer spin currents, resulting in the occurrence of ultrafast demagnetization \cite{turgut2013controlling} and magnetization enhancement \cite{rudolf2012ultrafast} in a separate FM layer. Furthermore, it has been found that excitation of NM metal can also trigger ultrafast demagnetization in the FM layer \cite{eschenlohr2013ultrafast}, but direct evidence for ultrafast magnetization enhancement induced by NM excitation remains elusive. A direct question would be whether an additional FM metal is required to achieve ultrafast magnetization enhancement. The search for the missing piece of the puzzle concerning interlayer spin transfer continues.

Currently, the physical mechanisms responsible for the enhancement of magnetization induced by ultrafast spin current remain controversial, and there has not been a consensus in the understanding of ultrafast hot electron interlayer transport. Following the discovery of laser-induced ultrafast magnetization enhancement, Turgut {\it et al.} investigated the contribution of spin transport and spin flip mechanisms in the Ni/Ru/Fe system \cite{turgut2013controlling}, where the nonlocal interlayer spin transport accounted for more than 2/3 of the Ni demagnetization contribution. When the demagnetization in the Ni layer reached 68$\%$, the Fe layer was found to have approximately 16$\%$ enhancement in its magnetization. Nonetheless, subsequent research conducted by Schellekens {\it et al.} using a novel all-optical method demonstrated that this unexpectedly strong magnetization enhancement in the Fe layer could not be reproduced, even though they also recognized the importance of the spin current transport \cite{schellekens2014exploring}. Furthermore, Eschenlohr {\it et al.} measured the element-resolved X-ray magnetic circular dichroism of a similar system but failed to find transient magnetization enhancement \cite{eschenlohr2017spin}. Instead, they observed a stronger demagnetization of the Fe layer in Ni/Ta/Fe compared with Ni/Ru/Fe, indicating that the spin scattering in the Ta layer suppressed the spin current from the Ni layer into the Fe layer. The authors argued that the sample thickness could be optimized for efficient utilization of ultrafast spin current, and further researches are needed to clarify the role of interfaces in the considered multilayer structure. More recently, Stamm {\it et al.} found that the demagnetization amplitude of the Ni and Fe layers had a difference of about 4.1$\%$ in the parallel and antiparallel configurations, which was attributed to the influence of interlayer spin transfer generated by the femtosecond laser pulse \cite{stamm2020xray}. So far, the interfaces in the multilayers have not been considered yet in the interpretation of experimental observations, although the interfacial properties were found to have significant influence on spin transport \cite{eschenlohr2017spin, seifert2022spintronic}. By systematically considering all the mechanisms for the ultrafast spin transport associated with the laser excited magnetization dynamics, we can manipulate the spin injection and dissipation in magnetic multilayers for the design of the ultrafast spintronic devices.

\begin{figure}[t]
  \centering
  \includegraphics[width=\columnwidth]{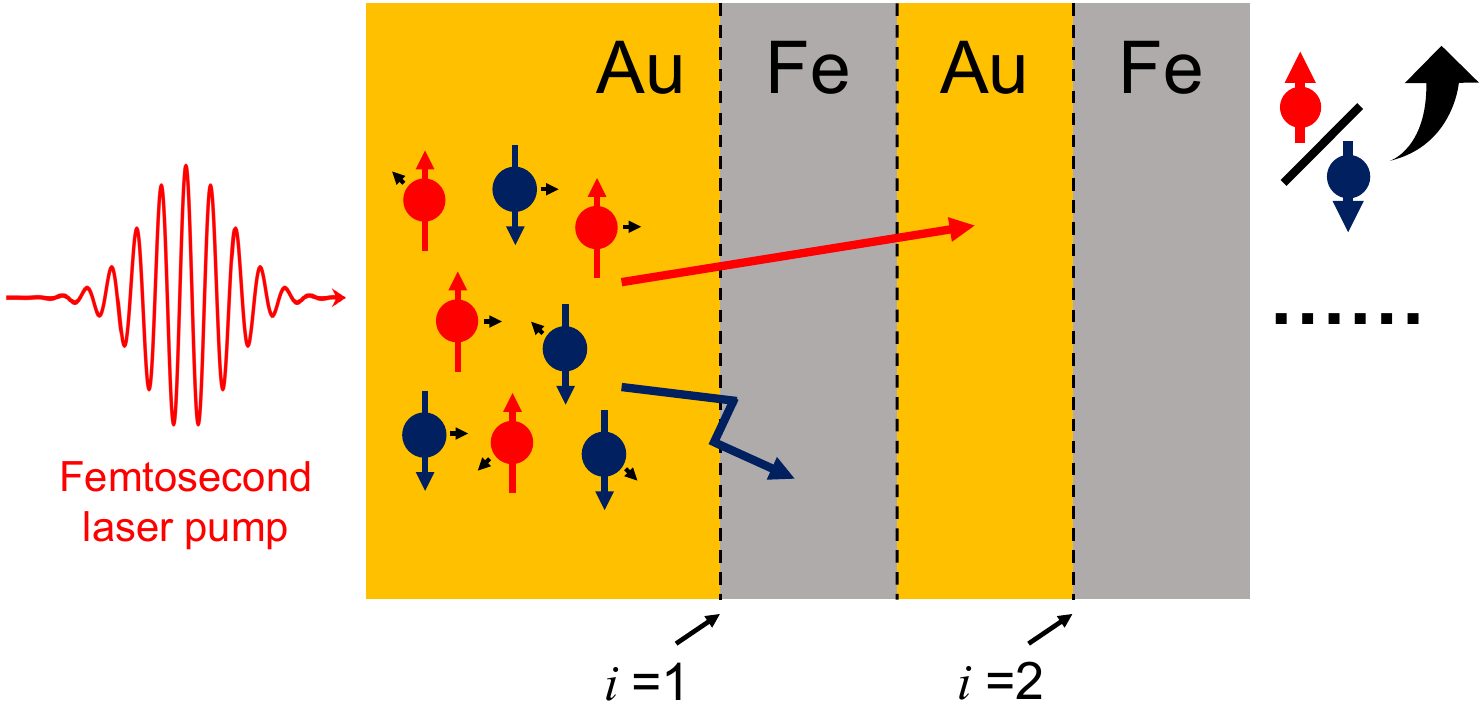}
  \caption{Schematic diagram of superdiffusive spin transport and spin injection in a [Au/Fe]$_n$ multilayer. A femtosecond laser pulse excites hot electrons in the first Au layer. With multiple spin filtering interface reflections and spin-dependent mobility experienced in the Fe layer, the resulting ultrafast spin current injected into the interior gradually increases its polarizability. Interface index $i$ denotes the particular Au/Fe interface with respect to the excited Au layer.}
  \label{f1}
\end{figure}

In this paper, we employ the theory of superdiffusive spin transport and conduct a systematic analysis of spin injection into a magnetic multilayer, which is excited by a femtosecond laser pulse as schematically illustrated in Fig.~\ref{f1}. The multilayer structure is primarily constructed by the repetition of Au/Fe bilayer units. In a single Au/Fe bilayer, the hot electrons excited in the Au layer effectively lead to the injection of spins into the Fe layer, where the magnetization enhancement or reduction may occur depending on the specific scattering at the Au/Fe interface. By increasing the repeated Au/Fe units, the multilayer gradually increases the polarization of injected spin current carried by laser-excited hot electrons, approaching to 100\% polarization. Both the bulk Fe layer and the Au/Fe interfaces serves as spin filters. The injected spin flux exhibits a nonmonotonic dependence on the number of repeated units and the maximum spin flux is found at the injection into the second Fe layer. The Au/Ni multilayers are studied for comparison. Even though the interfacial spin filtering effect is weak, the Au/Ni multilayer still generate highly spin polarized current efficiently.

The rest of this paper is organized as follows. The theoretical methods and computational details are briefly presented in Sec.~\ref{chap2}. The calculated results for spin injection in Au/FM multilayers are shown in Sec.~\ref{chap3}. Specifically, Sec.~\ref{chap3.1} focuses on the laser-excited magnetization enhancement or reduction in the Au/Fe bilayer, which is followed by the analysis of the ultrafast spin current injection in the multilayers in Sec.~\ref{chap3.2}. The optimal thickness for utilizing the ultrafast spin current is discussed in Sec.~\ref{chap3.3}. A comparative study for the Au/Ni multilayers is performed in Sec.~\ref{chap3.4} including the laser-induced magnetization dynamics and the associated ultrafast spin injection. Conclusions drawn in this study are given in Sec.~\ref{chap4}.

\section{Theoretical methods and computational details}\label{chap2}

The superdiffusive spin transport model successfully explained the experimental findings of ultrafast demagnetization in magnetic bilayers induced by femtosecond laser pulses \cite{battiato2010superdiffusive}. In this model, when a magnetic metal is excited by a laser pulse, a lot of nonequilibrium hot electrons are excited from the occupied states below the Fermi energy ($E_F$) to the energy bands above $E_F$. The properties of nonequilibrium electrons, including their lifetimes and group velocities, are dependent on both the band structure of the material and its spin in magnetic metals. In iron, for instance, a photon with an energy of 1.5~eV excites the majority-spin 3$d$ electrons to the unoccupied 4$s$ band, whereas the minority-spin 3$d$ electrons are excited to other unoccupied 3$d$ bands. As a consequence of photon excitation, the nonequilibrium electrons in Fe exhibit spin-dependent group velocities since 4$s$ electrons moving much faster than 3$d$ electrons. This feature leads to the emergence of a spin-polarized current carried by hot electrons. This model does not consider holes in the 3$d$ bands below $E_F$ due to their relatively low mobility\cite{ma2019plasmon}. Above $E_F$, hot electrons propagate with their own group velocity and, due to energy loss caused by scattering, recombine with holes below $E_F$. The propagation of hot electrons is influenced by scattering mechanisms such as phonons, impurities, and other electrons. Both elastic and inelastic scattering can occur, with the former conserving the electron energy but changing its momentum and the latter allowing the electrons to lose their energy. The released energy due to the inelastic scattering can be transferred to excite another electron to a higher energy state, leading to a cascade of electrons that contribute to the superdiffusive spin transport.

The laser-excited hot electrons may propagate across the metallic interface and enter the attached magnetic or nonmagnetic layer retaining their original energies. Nevertheless, at the interface of FM/NM or NM/FM, the spin-dependent interface resistance \cite{Xia:prb06} results in the unequal transmission probabilities of the opposite spins leading to the transfer of spin angular momentum between the adjacent materials. In this work, we explicitly calculate the spin injection efficiency in Au/FM multilayers after laser excitation in the nonmagnetic Au. The choice of utilizing an Au layer as the excitation layer stems from its superior infrared light absorptivity compared to ferromagnetic metals, combined with its notably long electron mean free path and spin decay length \cite{lu2021spin}. The details of the numerical techniques have been extensively documented in our previously published paper \cite{lu2020interface}.

Owing to the relatively larger size of the laser spot in comparison to the mean free path of the excited hot electrons, the transport can be simplified to a one-dimensional process along the interface normal of the metallic multilayers, specifically the $z$-axis. In the superdiffusive spin transport model, the key equation is as follows \cite{battiato2012theory}:
\be
\frac{\partial n_{\sigma}(E,z,t)}{\partial t} + \frac{n_{\sigma}(E,z,t)}{\tau_{\sigma}(E,z)} = \left( -\frac{\partial}{\partial z} \hat{\phi}+\hat{I} \right) S^{\rm eff}_{\sigma}(E,z,t),\label{eq:key}
\ee
where $n_{\sigma}(E,z,t)$ is the density of nonequilibrium hot electrons with spin $\sigma$ at energy $E$, position $z$ and time $t$. The lifetime of hot electrons with spin $\sigma$ at energy $E$ is represented by $\tau_{\sigma}(E,z)$, which is dependent on the particular material at the position $z$. Thus, the second term on the left-hand side of the equation describes the decay of nonequilibrium hot electrons. The effective source term $S_{\sigma}^{\text{eff}}(E,z,t)$ in Eq.\eqref{eq:key} includes both scattered and newly excited electrons. Operators $\hat{\phi}$ and $\hat{I}$ represent the electron flux and identity operators, respectively.

The flux operator $\hat{\phi}$ acting on the source term $S(z, t)$ includes the nonlocality in both space and time, i.e.,
\be
\hat{\phi} S(z,t) = \int^{+\infty}_{-\infty} dz_{0} \int^{t}_{-\infty} dt_{0}\ S(z_0,t_0)\phi(z,t|z_0,t_0).
\ee
Here the spin and energy indices are omitted for simplicity. The flux kernel $\phi(z, t|z_0, t_0)$ denotes the electron density flux at a given position $z$ and time $t$ due to an electron that was excited at $z_0$ and $t_0$. Because the superdiffusive spin transport equation \eqref{eq:key} is nonlocal, it has to be solved iteratively \cite{battiato2014treating}.

The time-dependent magnetization, which is the difference of the majority- and minority-spin electron density, is defined as follows:
\begin{eqnarray}
M(z, t) &=& \int dE \left[n_{\uparrow}(E, z, t) - n_{\downarrow}(E, z, t)\right] \nonumber\\
&&+n^{\rm occ}_{\uparrow}(z, t)-n^{\rm occ}_{\downarrow}(z, t). \label{eq:m}
\end{eqnarray}
Here, $n^{\rm occ}_{\sigma}(z, t)$ represents the electron density below the Fermi level with spin $\sigma$. This quantity is the summation of the electrons that were not excited and the electrons that dropped to a lower energy than $E_F$ due to inelastic scattering. The occupied electrons have relatively low mobility compared to the excited electrons above $E_F$, and therefore, their contribution to the spin current is neglected. Thus, the spin current density is defined as the difference between spin-up and spin-down electron flux densities,
\begin{eqnarray}
j_{s}(E,z,t) = &&\frac{\hbar}{2} [\Phi_{\rightarrow}^{\uparrow}\left(E,z,t\right)-\Phi_{\rightarrow}^{\downarrow}\left(E,z,t\right) \nonumber\\
&&-\Phi_{\leftarrow}^{\uparrow}\left(E,z,t\right)+\Phi_{\leftarrow}^{\downarrow}\left(E,z,t\right)].
\end{eqnarray}

In this work, we investigate the spin injection via hot electrons in the [Au/FM]$_n$ multilayers, where FM represents Fe or Ni, and $n=1,2,3,4$ is the number of repeated stacking units. To ensure that the laser excitation had an appropriate penetration depth, the thickness of the Au layer that is excited by the laser pulse is fixed as 10 nm, while the thickness of the remaining Au and FM metal layers was maintained as 5 nm (Fig.~\ref{f1}). In Sec.~\ref{chap3.3}, taking [Au/Fe]$_3$ as a typical example, we also searched for the optimal spin injection by varying the thickness of the non-excited metal layer. The excitation laser pulse is modeled using a Gaussian function with a wavelength of 780 nm (1.5~eV), which is commonly used in experiments, and a full width at half maximum (FWHM) of 60~fs. We set an equal number of spin-up and spin-down electrons to be excited by the laser pulse in the nonmagnetic Au, consistent with a previous calculation\cite{battiato2012theory}. In our simulation, the nonequilibrium hot electrons within the energy range of $E_F$ to $E_F+1.5$eV are discretized with an interval of 0.125eV. A spatial grid with a resolution of 0.5~nm and a time step of 1~fs are used. The determination of spin-dependent lifetimes and velocities of excited electrons was achieved by means of first-principles many-body calculations, as documented in literature~\cite{zhukov2005gw+, zhukov2006lifetimes}. The transition probability for electron scattering is set to be the same as previous works \cite{lu2020interface}.

The calculation being considered does not explicitly incorporate phonons, yet it implicitly includes phonon scattering in the elastic scattering transition probability, which results in the change in the direction of nonequilibrium hot electron motion \cite{battiato2012theory}. The relaxation times for electron-phonon and spin-phonon interactions are comparable to, or longer than, the total calculation time and are typically on the order of several picoseconds. Therefore, we have neglected electron thermalization due to these interactions, except when studying spin dynamics on a larger timescale.

\section{Results and discussions}\label{chap3}

Before presenting the detailed results, we define some particular terminologies to make the following presentation clear and consistent. Specifically, we use the term ``transparent interface'' to refer to an interface with zero reflectivity. In this case, the hot electrons arriving at the interface from either side can pass through without any reflection. For real interfaces, the energy and spin dependent reflectivity of hot electrons are explicitly calculated using first-principles scattering theory. Following our previous publication \cite{lu2020interface}, we consider two types of interfaces: clean and disordered. A ``clean'' interface is represented by a sharp planar boundary between two materials, while a ``disordered'' interface is characterized by atomic interdiffusion that leads to the formation of thin alloying layers at the interface. The interface resistance of FM/NM has been calculated in the energy range of $0-1.5$~eV above the Fermi level \cite{lu2020interface}. For the Fe/NM interface, the spin-up electrons have a higher transmission probability than the spin-down electrons in both incident directions, leading to a spin filtering effect. On the contrary, in the Ni/NM systems, the interfacial transmission probability of hot electrons is basically independent of spin. The calculated interfacial reflectivity is incorporated into the superdiffusive spin transport model to calculate the laser-induced magnetization dynamics and terahertz emission of the FM/NM bilayer \cite{lu2020interface}.

\subsection{The Au/Fe bilayer}\label{chap3.1}
\begin{figure}[t]
  \centering
  \includegraphics[width=\columnwidth]{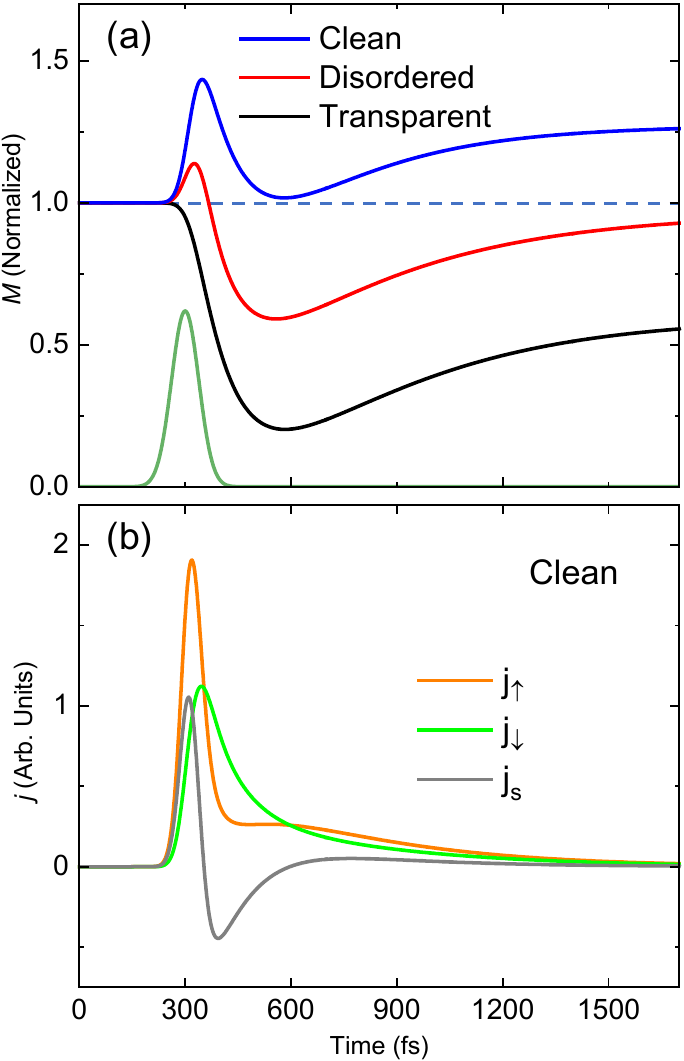}
  \caption{(a) Calculated magnetization dynamics of Fe after laser excitation in an Au/Fe bilayer with three distinct interface types: transparent (black line), clean (blue line), and disordered (red line). The light green line depicts the profile of the laser pulse employed. (b) The spin-up and down flows of hot electrons at the clean Au/Fe interface are represented by the orange and green lines, respectively. The difference between these currents is depicted as the gray line, which is known as the spin current.}
  \label{f2}
\end{figure}

The ultrafast spin dynamics in an Au/Fe bilayer system excited by a femtosecond laser pulse are calculated using the superdiffusive spin transport model, with the maximum pulse amplitude occurring at $t=300$~fs. The magnetization of Fe at the transparent, clean, and disordered interfaces, defined by Eq. (\ref{eq:m}), is plotted respectively in Fig.~\ref{f2}(a) as a function of time. The profile of the Gaussian laser pulse with a FWHM of 60~fs is also shown by the light green line for reference. It is quite unexpected that different treatments of interface reflectivity result in significantly different magnetization dynamics after the same laser excitation.

With the transparent interface, an ultrafast demagnetization process of Fe is observed, characterized by a rapid decrease in magnetization occurring within an approximately 300~fs after laser excitation, as shown by the black line in Fig.~\ref{f2}(a). Upon laser excitation, hot electrons are generated in Au, which possess a long diffusion length, enabling them to rapidly diffuse into the ferromagnetic layer, regardless of their spin orientation. However, in the Fe layer, the spin-up hot electrons have longer lifetimes and larger group velocities, which allows them to propagate for a relatively long distance. Thus, a considerable portion of these spin-up hot electrons flows back to Au after being reflected at the Fe/vacuum surface, resulting in a loss of net magnetization in the Fe layer. Meanwhile, the spin-down hot electrons injected from Au gradually decay in the Fe layer due to a much shorter lifetime. This demagnetization process is subsequently succeeded by a slow magnetic recovery process, which is attributed to the secondary injection of spin-up electrons.

With the real interface transmission, the scenario is remarkably different from the case with the transparent interface. With the clean interface, the magnetization intensity of Fe rapidly increases after the laser irradiation, reaching its maximum value at approximately 50 fs. Then the magnetization decays very quickly, approaching to its original value before the laser excitation. Subsequently, there is a slow and weak increase in magnetization starting from t=570 fs. This interesting ultrafast magnetization enhancement arises from the bi-directional spin filtering effect of hot electrons at the Au/Fe interface \cite{lu2020interface}. At the clean Au/Fe interface, the spin-up electrons have a higher transmission probability than spin-down electrons. After the laser excitation, a large number of spin-up hot electrons preferentially enter the ferromagnetic layer, leading to a quick increase in magnetization, which is in sharp contrast to the ``ultrafast demagnetization'' with the transparent interface. Because these spin-up hot electrons have high mobility, they are quickly reflected at the vacuum interface and flowing back to Au. This is the reason why the Fe layer loses its enhanced magnetization very soon. This process is similar to that in the bilayer system with the transparent interface, and therefore the minimum magnetization is found in the comparable time scale. Next, the spin-up hot electrons enter Fe again resulting in a small increase of magnetization. It is worth noting that the magnetization of the Fe layer is always higher than its equilibrium magnetization before excitation owing to the strong spin filtering effect.

To gain a better understanding, we calculate the current density $j_{\sigma}$ at the clean Au/Fe interface, which is carried by the hot electrons with spin $\sigma$. The calculated current density $j_{\sigma}$ as a function of time is shown in Fig.~\ref{f2}(b). Here positive values correspond to the current density flowing from Au towards Fe. After laser excitation, the spin-up hot electron current $j_{\uparrow}$ (the orange line) reaches its maximum value very quickly, and then decays due to the backflow after being reflected by the Fe/vacuum surface. It also shows a very weak and broad peak around t$\sim$600~fs corresponding to spin-up hot electrons entering Fe for the second time. The spin-down hot electron current (the green line) has a lower maximum in its intensity because it is strongly reflected at the Au/Fe interface. There is only a single broad peak in $j_{\downarrow}$, indicating that most of spin-down hot electrons only enter Fe once. Consequently, the spin current across the Au/Fe interface exhibits the shape with a peak followed by a valley, as shown by the gray line in Fig.~\ref{f2}(b).

Prior research has demonstrated that the spin filtering effect at disordered interfaces is weaker than that for the clean interfaces \cite{lu2020interface}. Therefore, the magnetization enhancement of the Fe layer with the disordered Au/Fe interface still occurs but it is weaker than that with the clean interface. The significant difference is observed when the spin-up hot electrons flow back to the Au layer: the magnetization of the Fe layer becomes smaller than its original equilibrium value indicating that a demagnetization is followed by the ultrafast enhancement. The minor magnetization recovery is also seen in Fe due to the re-entry of spin-up hot electrons. Thus, we can conclude based on the data shown in Fig.~\ref{f2} that even though the spin filtering effect is lowered down by the interface disorder, it can still induce the ultrafast enhancement of magnetization. Consequently, the spin filtering effect must be properly considered when the laser-excited magnetization dynamics is investigated. In a recent experiment, Jiang {\it et al.} observed the ultrafast magnetization enhancement and demagnetization of the ferromagnetic layer in the MnGa/GaAs bilayer \cite{jiang2022ultrafast}, which is in agreement with the magnetization curve of disordered interface in Fig.~\ref{f2}. In addition, the difference in magnetization increase and decrease after the laser excitation observed in experiments may be attributed to the different influences of the interfaces.

\subsection{Ultrafast spin current in Au/Fe multilayers}\label{chap3.2}

Based on the aforementioned phenomena in the Au/Fe bilayer, we propose to employ a multilayer structure with the repeated stacking units [Au/Fe]$_n$, which is capable of generating highly spin-polarized electron currents to facilitate the control of future ultrafast spintronic devices. In such magnetic multilayers, the repeated Au/Fe interfaces can increase the spin polarization of the hot electron current via multiple spin filtering scattering. Moreover, the strong reflection at the Fe/vacuum interface in the Au/Fe bilayer, which effectively increases the loss of magnetization, can be reduced.

To verify the above proposal, we now perform the calculations of spin injection for [Au/Fe]$_n$ multilayer with the number of stacking units $n=$1, 2, 3, and 4. Explicitly, we define the total spin polarization $P$ of the transient current entering the interface as
\begin{equation}
P= \frac{\int (j_{\uparrow}-j_{\downarrow}) dt}{\int (j_{\uparrow}+j_{\downarrow}) dt} \times 100\% \label{eq:p}
\end{equation}
where $j_{\uparrow}$ ($j_{\downarrow}$) is the current density carried by spin-up (spin-down) hot electrons.

\begin{figure*}[t]
  \centering
  \includegraphics[width=2\columnwidth]{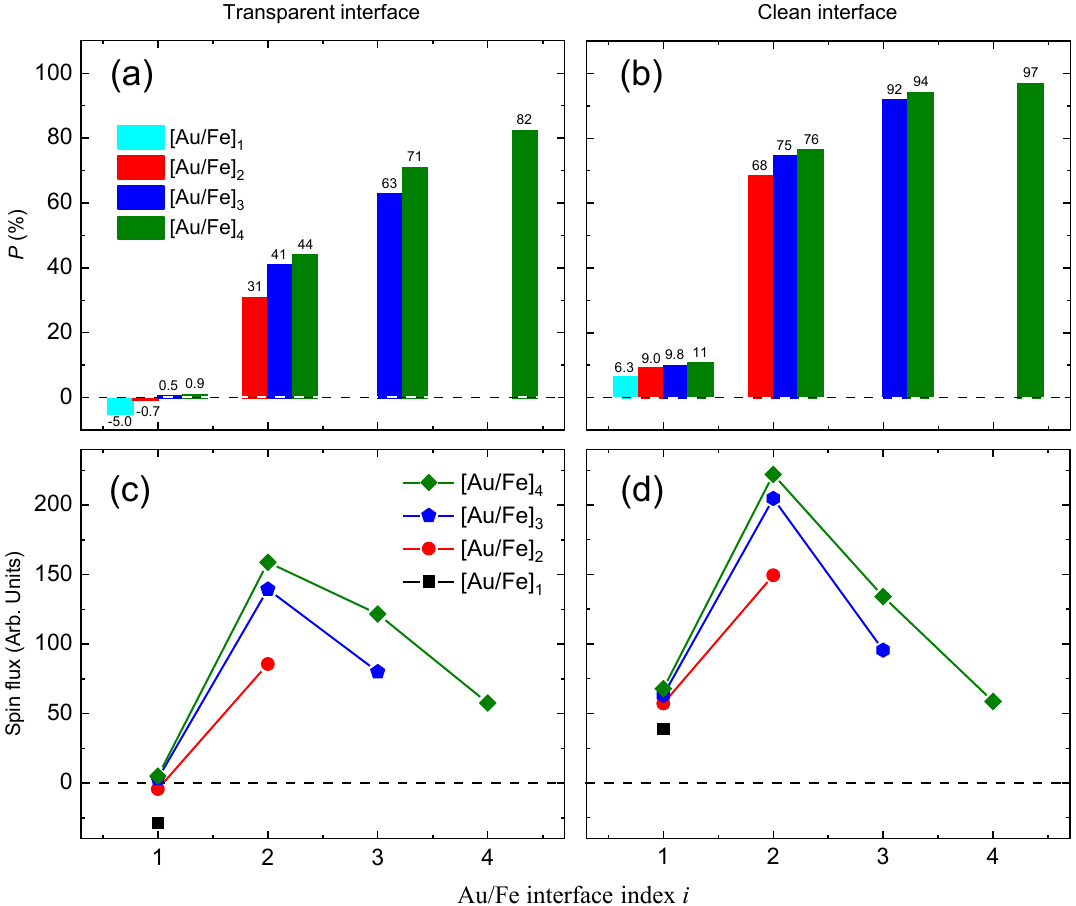}
  \caption{Calculated spin polarization and spin flux at each Au/Fe interface for the transparent and clean scenarios. Specifically, (a) and (b) show the spin polarization at each interface for the transparent and clean cases, respectively, while (c) and (d) display the spin flux at each Au/Fe interface. The indices of the Au/Fe interfaces are numerated from the excited Au layer, which are schematically illustrated in Fig.~\ref{f1}.}
  \label{f3}
\end{figure*}
To clearly illustrate the effect of interface reflection on the spin injection, we perform separate calculations with the transparent and clean Au/Fe interfaces, respectively. For the stacking number $n$, there are $n$ Au/Fe interfaces, which are numbered as $i=1,\cdots,n$ from the excited Au layer to the farthest Fe layer (see Fig.~\ref{f1}). The calculated spin polarization at the $i$-th interface in the [Au/Fe]$_n$ multilayer with the transparent interfaces is shown in Fig.~\ref{f3}(a). For the Au/Fe bilayer ($n=1$), the total spin polarization is negative, which is consistent with the ultrafast demagnetization after the laser excitation calculated in Fig.~\ref{f2}(a). At $n>1$, the spin polarization across the first interface gradually increases with increasing $n$ and becomes positive for $n\ge3$. This is because the backflow of spin-up hot electrons due to the reflection at the final vacuum interface is significantly reduced with larger $n$.

At the second interface ($i=2$) for the multilayers ($n\ge 2$), the spin polarization is largely enhanced to $30\sim40\%$. Although the transparent interface does not have the spin filtering effect, the spin polarization increases to $60\sim70\%$ at $i=3$ and up to $82\%$ at $i=4$. This is because $j_{\downarrow}$ carried by the spin-down hot electrons decays very quickly with increasing the stacking units $n$. Therefore, the Fe layers serves as the natural spin filter owing to the low velocity and short lifetime of the spin-down hot electrons.

The clean Au/Fe interfaces provide additional spin filtering effect and thus the spin polarization shown in Fig.~\ref{f3}(b) are larger than those in Fig.~\ref{f3}(a) for every interface and stacking number $n$. $P$ is always positive at the first interface and it becomes as large as 97\% at the last interface in [Au/Fe]$_4$. The latter is approaching a fully polarized spin current. These results suggest the synergistic effect between the interfacial spin filtering and the bulk spin filtering by stacking more units. For both transparent and clean interfaces, the spin polarization at the $i$-th interface is comparable for different numbers $n$ of the stacking units.

For the purpose of utilizing the ultrafast spin current, not only does the spin polarization play an important role, but so does the intensity of the spin current. To quantitatively evaluate the latter one, we calculate the spin flux across each Au/Fe interface defined by
\begin{equation}
F= \int (j_{\uparrow}-j_{\downarrow}) dt, \label{eq:A}
\end{equation}
which is merely the numerator of Eq.~\eqref{eq:p}. The calculated spin flux $F$ at each interface are shown in Fig.~\ref{f3}(c) and (d) for the transparent and clean interfaces, respectively. The spin flux is relatively small at the first Au/Fe interface, i.e. $i=1$. With the transparent interface, the total flux $F$ at the first interface is negative for $n=1$ and is close to zero for $n\ge2$. At $i=2$, the spin flux reaches its maximum for $n\ge2$. If the stacking units increases further, the decay of hot electrons reduces the total spin flux despite of the increase in the spin polarization. For the clean interface, the scenario is very similar except for a global increase in the spin flux. This is because of the additional spin filtering effect at the clean interfaces. It is worth noting that, at the same position of $i=2$, the spin flux increases with increasing $n$ for both transparent and clean interfaces, indicating that the effect of the reflection at the final vacuum interface is gradually diminished with large $n$. Considering both the two factors, spin polarization and spin current amplitude, we conclude that a multilayer structure with 2 or 3 stacking layers is optimal for utilizing the ultrafast spin current in the future spintronic devices.

\subsection{Thickness optimization}\label{chap3.3}
In the above calculations, we use a 10-nm-thick Au layer as the excitation layer, and the other alternately stacked Fe and Au layers all have the thickness of 5 nm. Due to the high sensitivity of hot electron attenuation to the multilayers limited by scattering, we now proceed to study the influence of film thickness on the spin injection to find the optimal efficiency for application.

\begin{figure}[t]
  \centering
  \includegraphics[width=\columnwidth]{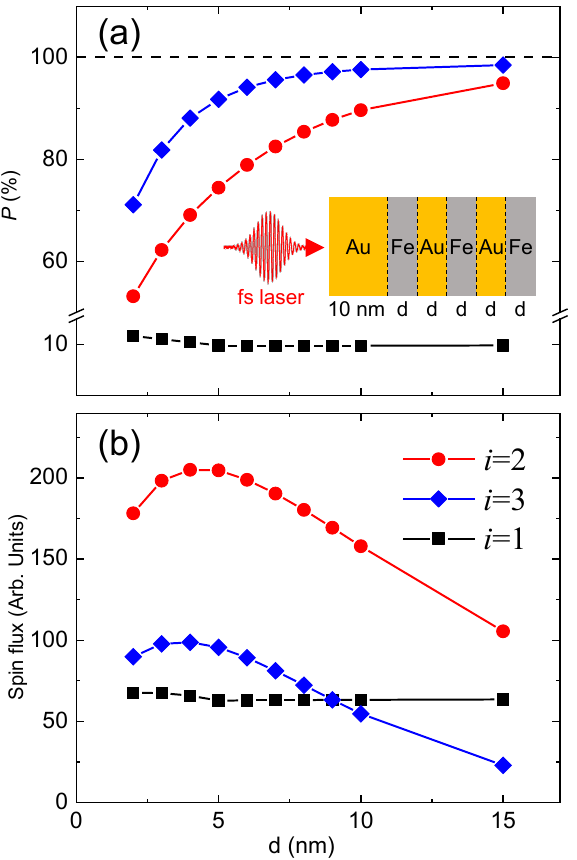}
  \caption{Calculated spin polarization (a) and spin flux (b) at each Au/Fe interfaces in the [Au/Fe]$_3$ multilayer films with various thicknesses for the nonexcited layers.}
  \label{f4}
\end{figure}

Based on the results of the previous section, we focus on the [Au/Fe]$_3$ system, in which the spin currents across the Au/Fe interfaces exhibit relatively large intensity and high spin polarization. With the fixed thickness of 10 nm for the excited Au layer, we simultaneously vary the thickness $d$ of all the other Au and Fe layers and calculate the spin polarization $P$ and spin flux $F$ at every Au/Fe interfaces, as plotted in Fig.~\ref{f4}(a) and (b), respectively. The clean Au/Fe interfaces are employed in this calculation.

The calculated polarization and spin flux injected through the first Au/Fe interface are nearly independent of the thickness $d$, as shown by the black squares in Fig.~\ref{f4}(a) and (b). This is because they are basically determined by the direct excitation of the 10-nm-thick Au layer. Only at very small $d$, the strong backflow slightly influences the calculated results. For the other two interfaces, i.e. $i=2$ and $3$, the calculated spin polarization increases monotonically with increasing $d$; see Fig.~\ref{f4}(a). The injected current into the third Au/Fe interface approaches the full polarization at $d>7$ nm. The spin polarization of the current injected into the second interface is already 95\% at $d=15$ nm. Such a large spin polarization manifests the significant spin filtering effect of bulk Fe.

Different from the monotonic thickness-dependence of the spin polarization, the spin flux injected through the second and the third Au/Fe interfaces exhibit a clear nonmonotonic dependence on $d$, as plotted by the red circles and blue diamonds in Fig.~\ref{f4}(b). At very small $d$, the excited hot electrons move through multiple metallic layers. When the thickness of each layer $d$ decreases, the reflections of hot electrons at interfaces are strongly enhanced. In addition, there exist many hot electrons reflected by the Fe/vacuum interface at small $d$. Both these two factors result in the decrease of the injected spin flux at the second and third Au/Fe interfaces. At a large thickness $d$, on the contrary, the hot electrons can propagate a relatively long distance and the bulk scattering becomes the main dissipation mechanism. Thus, increasing $d$ results in a decreasing spin flux arriving at the second and the third Au/Fe interfaces. In particular, the spin flux injected into the third Au/Fe interface is much smaller than that into the second interface, in agreement with blue symbols shown in Fig.~\ref{f3}(d). Except for the spin flux injected through the first Au/Fe interface, all the others are expected to decay to zero at enough large $d$. From the calculated spin polarization and spin flux shown in Fig.~\ref{f4}, we conclude that the optimal spin injection shall be achieved within a thickness range of $d = 4-5$ nm.

\subsection{The case of Au/Ni multilayers}\label{chap3.4}

\begin{figure}[t]
  \centering
  \includegraphics[width=\columnwidth]{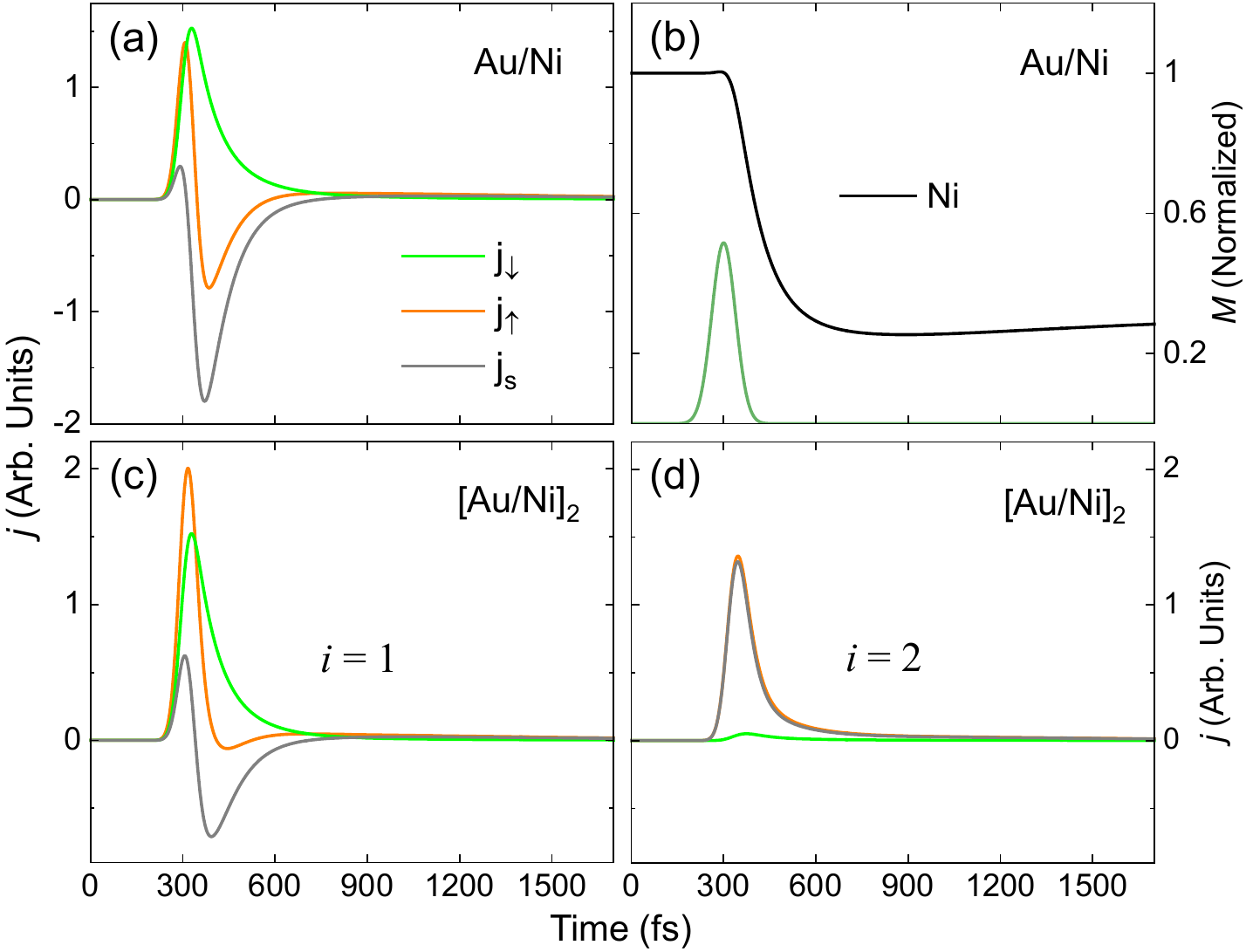}
  \caption{Calculated currents carried by the spin-up and spin-down hot electrons, and their difference at the clean Au/Ni interface in the Au/Ni bilayer (a) and in the [Au/Ni]$_2$ multilayer (c), (d). For the [Au/Ni]$_2$ multilayer, the calculated results at the first ($i=1$) and the second ($i=2$) Au/Ni interface are plotted in (c) and (d), respectively. (b) Calculated magnetization of Ni in the Au/Ni bilayer as a function of time. The light green line in (b) illustrates the profile of the laser pulse to excite the Au layer.}
  \label{f5}
\end{figure}

We then replace the Fe layers by Ni and study the [Au/Ni]$_n$ ($n=1\sim 4$) multilayer system. We calculate the superdiffusive spin transport with clean Au/Ni interfaces, although the interfacial spin filtering effect is relatively weak compared with that at the Au/Fe interface \cite{lu2020interface}. As in the [Au/Fe]$_n$ multilayers, the excitation layer of Au is fixed with its thickness of 10 nm and all the other layers have the same thickness of 5 nm. We start from the Au/Ni bilayer, in which the calculated hot-electron currents across the interface are plotted in Fig.~\ref{f5}(a) as a function of time. The current carried by the spin-up hot electrons increases very quickly after the excitation, as shown by the sharp peak in the orange line, which is followed by a broader valley due to the backflow of spin-up hot electrons. In fact, because the 3$d$ bands in Ni is completely occupied for the spin-up states, the spin-up hot electrons in Ni are 4$s$ states with large group velocities and long lifetime. Therefore, many spin-up electrons are reflected at the Ni/vacuum interface and flow back to Au resulting in the valley in the orange curve. Due to the short lifetime of the spin-down hot electrons, they are mostly scattered and decay inside Ni. So there is negligible backflow of spin-down hot electrons, as shown by the green line in Fig.~\ref{f5}(a). Consequently the total spin injected into the Ni layer is negative indicating the loss of magnetization after the laser excitation. This effective demagnetization is seen in the calculated magnetization of Ni as a function of time, shown in Fig.~\ref{f5}(b). At $t=300$~fs, there is a very weak magnetization enhancement corresponding to the peak of $j_s$, which is followed by the strong demagnetization in the next 1 ps. Experimentally, the magnetization enhancement is hardly detected in Ni because of the strikingly different characteristics between spin-up and spin-down hot electrons.

By increasing the number of stacking layers ($n=2$), we calculate the spin-dependent current across the two Au/Ni interfaces, as plotted in Fig.~\ref{f5}(c) and (d), respectively. At the first Au/Ni interface ($i=1$), the spin-dependent currents carried by the hot electrons are very similar to those in Fig.~\ref{f5}(a), except for the significantly reduced backflow (the negative peak) of the spin-up current $j_{\uparrow}$. This is because many spin-up hot electrons are injected into the second repeated Au/Ni unit and the reflection at the Ni/vacuum interface is suppressed. Figure~\ref{f5}(d) shows the extracted hot electron currents at the second Au/Ni interface ($i=2$), where the spin-down hot electrons become negligible due to the strong scattering in the first Ni layer. Only the spin-up electrons survive at the second Au/Ni interface resulting in a very high spin polarization.

\begin{figure}[t]
  \centering
  \includegraphics[width=\columnwidth]{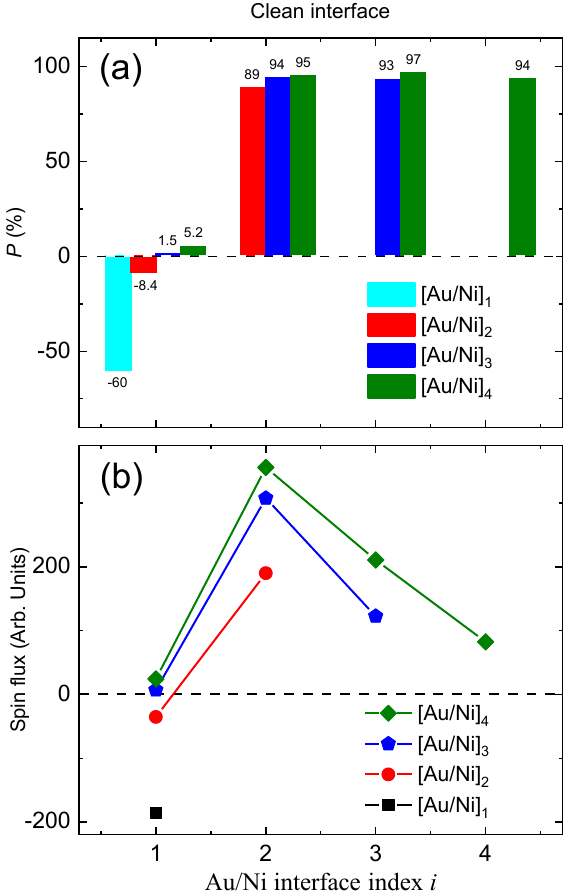}
  \caption{Calculated spin polarization (a) and spin flux (b) at each Au/Ni interface in the [Au/Ni]$_n$ multilayers. The reflectivity of the clean Au/Ni interface is employed in the calculation. The indices $i$ of the Au/Ni interfaces are numerated from the excited Au layer.}
  \label{f6}
\end{figure}
The calculated spin polarization and spin flux at each Au/Ni interface in the [Au/Ni]$_n$ multilayers are summarized in Fig.~\ref{f6}. In the bilayer, the spin polarization injected into Ni is $-60\%$ due to the large backflow of spin-up electrons shown in Fig.~\ref{f5}(a). With increasing in the number of stacking [Au/Ni] units, the spin polarization at the $i=1$ interface rapidly increases and becomes positive at $n=3$ owing to the reduction of the backflow of spin-up hot electrons. Surprisingly, at the second interface ($i=2$), the spin polarization is significantly enhanced to approximately $90\%$, which is independent of $n$. It again suggests the very strong spin filtering effect of the bulk Ni, as indicated by Fig.~\ref{f5}(d). After the electrons passing through more Ni layers, the spin polarization is always kept around 95\%.

Figure~\ref{f6}(b) shows the spin flux $F$ injected into the $i$-th Au/Ni interface in the [Au/Ni]$_n$ multilayers. At $n=1$, the large negative spin flux is related to the strong demagnetization shown in Fig.~\ref{f5}(b). For the other three multilayers, i.e. $n=2\sim4$, the spin flux reaches the maximum positive values at the second ($i=2$) Au/Ni interface. This scenario is very similar as in the [Au/Fe]$_n$ multilayers shown in Fig.~\ref{f3}(d). Further increasing the number of stacking units, the spin flux injected into the same $i$-th interface increases, highlighting the influence of the reflection at the final Ni/vacuum interface. Comparing to the [Au/Fe]$_n$ multilayers, the high spin polarization can be achieved with fewer stacking units, offering an advantage of [Au/Ni]$_n$ multilayers in future applications.

\section{Conclusions}\label{chap4}

We have systematically investigated the spin currents carried by hot electrons in [Au/FM]$_n$ (FM = Fe and Ni) multilayers with $n=1\sim4$ using the superdiffusive spin transport model. In the Au/Fe bilayer, hot electrons excited in Au induce ultrafast magnetization reduction or enhancement in the Fe layer depending on the specific interface reflection conditions. In comparison, the magnetization enhancement in the Au/Ni bilayer is very weak and can be hardly detected. This is because (1) the spin filtering effect is very weak at the Au/Ni interface, and (2) the spin-up electrons in Ni have much larger velocity and longer lifetime than the spin-down electrons. Our results explain the different experimental reports in literature and clarify the different features between Fe and Ni.

For the multilayers [Au/Fe]$_n$, the spin polarization of current monotonically increases and eventually approaches 100\%. The high spin polarization results from the spin filtering effect at the interfaces as well as in the bulk Fe. The latter is attributed to the distinctive group velocities and lifetimes between the spin-up and spin-down hot electrons. In ferromagnetic Ni, the bulk spin filtering effect is even more significant than that in Fe since the 3$d$ bands of the majority spin are completely occupied and only 4$s$ states are present in the laser excited hot electrons. Moreover, the intensity of spin flux injected into the Au/Fe interface exhibits a nonmonotonic dependence on the number of repeated stacking units in the multilayer, which can be attributed to the competition between hot electron backflow and energy dissipation during the dynamics of hot electrons. Our findings provide an effective method for generating a highly spin polarization current by an ultrafast laser pulse for the development of future spintronics devices.

\acknowledgements
This work at Shanxi Normal University is supported by the National Key R$\&$D Program of China (Grants No. 2022YFB3505301), the National Natural Science Foundation of China (NSFC, Grants No. 12174237), China Postdoctoral Science Foundation (Grants No. 2023M732150), and the Basic Research Plan of Shanxi Province (Grant No. 202203021212393 and No. 202203021222219). Z. Y. acknowledges the financial support by the NSFC (Grant No. 12174028).


\begin{thebibliography}{}

\bibitem{jhang2021challenges}
C.-J. Jhang, C.-X. Xue, J.-M. Hung, F.-C. Chang, and M.-F. Chang,
Challenges and trends of SRAM-based computing-in-memory for AI edge devices,
\href{https://doi.org/10.1109/TCSI.2021.3064189}{IEEE T. Circuits-I {\bf 68}, 1773 (2021).}

\bibitem{zou2021breaking}
X. Zou, S. Xu, X. Chen, L. Yan, and Y. Han,
Breaking the von Neumann bottleneck: architecture-level processing-in-memory technology,
\href{https://doi.org/10.1007/s11432-020-3227-1}{Sci. China Inform. Sci. {\bf 64}, 160404 (2021).}

\bibitem{fouda2022in}
M. E. Fouda, H. E. Yant{\i}r, A. M. Eltawil, and F. Kurdahi,
In-memory associative processors: tutorial, potential, and challenges,
\href{https://doi.org/10.1109/TCSII.2022.3170468}{IEEE T. Circuits-II {\bf 69}, 2641 (2022).}

\bibitem{wolf2001spintronics}
S. A. Wolf, D. D. Awschalom, R. A. Buhrman, J. M. Daughton, S. von Moln\'ar,  M. L. Roukes, A. Y. Chtchelkanova, and D. M. Treger,
Spintronics: a spin-based electronics vision for the future,
\href{https://doi.org/10.1126/science.1065389}{Science {\bf 294}, 1488 (2001).}

\bibitem{chappert2007the}
C. Chappert, A. Fert, and F. N. Van Dau,
The emergence of spin electronics in data storage,
\href{https://doi.org/10.1038/nmat2024}{Nat. Mater. {\bf 6}, 813 (2007).}

\bibitem{grollier2020neuromorphic}
J. Grollier, D. Querlioz, K. Y. Camsari, K. Everschor-Sitte, S. Fukami, and M. D. Stiles,
Neuromorphic spintronics,
\href{https://doi.org/10.1038/s41928-019-0360-9}{Nat. Electron. {\bf 3}, 360 (2020).}

\bibitem{hayakawa2021nanosecond}
K. Hayakawa, S. Kanai, T. Funatsu, J. Igarashi, B. Jinnai, W. A. Borders, H. Ohno, and S. Fukami
Nanosecond random telegraph noise in in-plane magnetic tunnel junctions
\href{https://doi.org/10.1103/PhysRevLett.126.117202}{Phys. Rev. Lett. {\bf 126}, 117202 (2021).}

\bibitem{beaurepaire1996ultrafast}
E. Beaurepaire, J.-C. Merle, A. Daunois, and J.-Y. Bigot,
Ultrafast spin dynamics in ferromagnetic nickel,
\href{https://doi.org/10.1103/PhysRevLett.76.4250}{Phys. Rev. Lett. {\bf 76}, 4250 (1996).}

\bibitem{kirilyuk2010ultrafast}
A. Kirilyuk, A. V. Kimel, and Th. Rasing,
Ultrafast optical manipulation of magnetic order,
\href{https://doi.org/10.1103/RevModPhys.82.2731}{Rev. Mod. Phys. {\bf 82}, 2731 (2010).}

\bibitem{lu2022progress}
W.-T. Lu and Z. Yuan,
Progress in ultrafast spintronics research (in Chinese),
\href{https://doi.org/10.1360/SSPMA-2021-0350}{Sci. Sin.-Phys. Mech. Astron. {\bf 52}, 270007 (2022).}

\bibitem{zhang2000laser}
G. P. Zhang and W. H\"ubner,
Laser-induced ultrafast demagnetization in ferromagnetic metals,
\href{https://doi.org/10.1103/PhysRevLett.85.3025}{Phys. Rev. Lett. {\bf 85}, 3025 (2000).}

\bibitem{koopmans2005unifying}
B. Koopmans, J. J. M. Ruigrok, F. Dalla Longa, and W. J. M. de Jonge,
Unifying ultrafast magnetization dynamics,
\href{https://doi.org/10.1103/PhysRevLett.95.267207}{Phys. Rev. Lett. {\bf 95}, 267207 (2005).}

\bibitem{carpene2008dynamics}
E. Carpene, E. Mancini, C. Dallera, M. Brenna, E. Puppin, and S. De Silvestri,
Dynamics of electron-magnon interaction and ultrafast demagnetization in thin iron films,
\href{https://doi.org/10.1103/PhysRevB.78.174422}{Phys. Rev. B {\bf 78}, 174422 (2008).}

\bibitem{krauss2009ultrafast}
M. Krau\ss, T. Roth, S. Alebrand, D. Steil, M. Cinchetti, M. Aeschlimann, and H. C. Schneider,
Ultrafast demagnetization of ferromagnetic transition metals: the role of the Coulomb interaction,
\href{https://doi.org/10.1103/PhysRevB.80.180407}{Phys. Rev. B {\bf 80}, 180407(R) (2009).}

\bibitem{bigot2009coherent}
J.-Y. Bigot, M. Vomir, and E. Beaurepaire,
Coherent ultrafast magnetism induced by femtosecond laser pulses,
\href{https://doi.org/10.1038/nphys1285}{Nat. Phys. {\bf 5}, 515 (2009).}

\bibitem{qaiumzadeh2013manipulation}
A. Qaiumzadeh, G. E. W. Bauer, and A. Brataas,
Manipulation of ferromagnets via the spin-selective optical Stark effect,
\href{https://dx.doi.org/10.1103/PhysRevB.88.064416}{Phys. Rev. B {\bf 88}, 064416 (2013).}

\bibitem{tveten2015electron}
E. G. Tveten, A. Brataas, and Y. Tserkovnyak,
Electron-magnon scattering in magnetic heterostructures far out of equilibrium,
\href{https://doi.org/10.1103/PhysRevB.92.180412}{Phys. Rev. B {\bf 92}, 180412(R) (2015).}

\bibitem{freimuth2021laser}
F. Freimuth, S. Bl\"ugel, and Y. Mokrousov,
Laser-induced torques in spin spirals,
\href{https://doi.org/10.1103/PhysRevB.103.054403}{Phys. Rev. B {\bf 103}, 054403 (2021).}

\bibitem{koopmans2010explaining}
B. Koopmans, G. Malinowski, F. Dalla Longa, D. Steiauf, M. F\"ahnle, T. Roth, M. Cinchetti, and M. Aeschlimann,
Explaining the paradoxical diversity of ultrafast laser-induced demagnetization,
\href{https://doi.org/10.1038/nmat2593}{Nat. Mater. {\bf 9}, 259 (2010).}

\bibitem{malinowski2008control}
G. Malinowski, F. Dalla Longa, J. H. H. Rietjens, P. V. Paluskar, R. Huijink, H. J. M. Swagten, and B. Koopmans,
Control of speed and efficiency of ultrafast demagnetization by direct transfer of spin angular momentum,
\href{https://doi.org/10.1038/nphys1092}{Nat. Phys. {\bf 4}, 855 (2008).}

\bibitem{battiato2010superdiffusive}
M. Battiato, K. Carva, and P. M. Oppeneer,
Superdiffusive spin transport as a mechanism of ultrafast demagnetization,
\href{https://doi.org/10.1103/PhysRevLett.105.027203}{Phys. Rev. Lett. {\bf 105}, 027203 (2010).}

\bibitem{eschenlohr2013ultrafast}
A. Eschenlohr, M. Battiato, P. Maldonado, N. Pontius, T. Kachel, K. Holldack, R. Mitzner, A. F\"ohlisch, P. M. Oppeneer, and C. Stamm,
Ultrafast spin transport as key to femtosecond demagnetization,
\href{https://doi.org/10.1038/nmat3546}{Nat. Mater. {\bf 12}, 332 (2013).}

\bibitem{kampfrath2013terahertz}
T. Kampfrath, M. Battiato, P. Maldonado, G. Eilers, J. N\"otzold, S. M\"ahrlein, V. Zbarsky, F. Freimuth, Y. Mokrousov, S. Bl\"ugel, M. Wolf, I. Radu, P. M. Oppeneer, and M. M\"unzenberg,
Terahertz spin current pulses controlled by magnetic heterostructures,
\href{https://doi.org/10.1038/nnano.2013.43}{Nat. Nanotech. {\bf 8}, 256 (2013).}

\bibitem{liu2021strain}
C. Q. Liu , W.-T. Lu, Z. X. Wei, Y. F. Miao, P. Wang, H. Xia, Y. P. Liu , F. L. Zeng, J. R. Zhang, C. Zhou, H. B. Zhao, Y. Z. Wu, Z. Yuan , and J. Qi,
Strain-induced anisotropic terahertz emission from a Fe(211)/Pt(110) bilayer,
\href{http://dx.doi.org/10.1103/PhysRevApplied.15.044022}{Phys. Rev. Appl. {\bf 15}, 044022 (2021).}

\bibitem{rudolf2012ultrafast}
D. Rudolf, C. La-O-Vorakiat, M. Battiato, R. Adam, J. M. Shaw, E. Turgut, P. Maldonado, S. Mathias, P. Grychtol, H. T. Nembach, T. J. Silva, M. Aeschlimann, H. C. Kapteyn, M. M. Murnane, C. M. Schneider, and P. M. Oppeneer,
Ultrafast magnetization enhancement in metallic multilayers driven by superdiffusive spin current,
\href{https://doi.org/10.1038/ncomms2029}{Nat. Commun. {\bf 3}, 1037 (2012).}

\bibitem{he2013ultrafast}
W. He, T. Zhu, X.-Q. Zhang, H.-T. Yang, and Z.-H. Cheng,
Ultrafast demagnetization enhancement in CoFeB/MgO/CoFeB magnetic tunneling junction driven by spin tunneling current,
\href{https://doi.org/10.1038/srep02883}{Sci. Rep. {\bf 3}, 2883 (2013).}

\bibitem{ji2023ultrafast}
B. Ji, Z. Jin, G. Wu, J. Li, C. Wan, X. Han, Z. Zhang, G. Ma, Y. Peng, and Y. Zhu,
Ultrafast laser-induced magneto-optical response of CoFeB/MgO/CoFeB magnetic tunneling junction,
\href{https://doi.org/10.1063/5.0141071}{Appl. Phys. Lett. {\bf 122}, 111104 (2023).}

\bibitem{turgut2013controlling}
E. Turgut, C. La-o-vorakiat, J. M. Shaw, P. Grychtol, H. T. Nembach, D. Rudolf, R. Adam, M. Aeschlimann, C. M. Schneider, T. J. Silva, M. M. Murnane, H. C. Kapteyn, and S. Mathias,
Controlling the competition between optically induced ultrafast spin-flip scattering and spin transport in magnetic multilayers,
\href{https://doi.org/10.1103/PhysRevLett.110.197201}{Phys. Rev. Lett. {\bf 110}, 197201 (2013).}

\bibitem{schellekens2014exploring}
A.J. Schellekens, N. de Vries, J. Lucassen, and B. Koopmans,
Exploring laser-induced interlayer spin transfer by an all-optical method,
\href{https://doi.org/10.1103/PhysRevB.90.104429}{Phys. Rev. B {\bf 90}, 104429 (2014).}

\bibitem{eschenlohr2017spin}
A. Eschenlohr, L. Persichetti, T. Kachel, M. Gabureac, P. Gambardella, and C. Stamm,
Spin currents during ultrafast demagnetization of ferromagnetic bilayers,
\href{https://doi.org/10.1088/1361-648X/aa7dd3}{J. Phys.: Condens. Matter {\bf 29}, 384002 (2017).}

\bibitem{stamm2020xray}
C. Stamm, C. Murer, M. S. W\"ornle, Y. Acremann, R. Gort, S. D\"aster, A. H. Reid, D. J. Higley, S. F. Wandel, W. F. Schlotter, and P. Gambardella,
X-ray detection of ultrashort spin current pulses in synthetic antiferromagnets,
\href{https://doi.org/10.1063/5.0006095}{J. Appl. Phys. {\bf 127}, 223902 (2020).}

\bibitem{seifert2022spintronic}
T. S. Seifert, L. Cheng, Z. Wei, T. Kampfrath, and J. Qi,
Spintronic sources of ultrashort terahertz electromagnetic pulses,
\href{https://doi.org/10.1063/5.0080357}{Appl. Phys. Lett. {\bf 120}, 180401 (2022).}

\bibitem{ma2019plasmon}
J. Ma, and S. Gao,
Plasmon-induced electron-hole separation at the Ag/TiO$_2$(110) interface,
\href{https://doi.org/10.1021/acsnano.9b03555}{ACS nano {\bf 13}, 13658 (2019).}

\bibitem{Xia:prb06}
K. Xia, M. Zwierzycki, M. Talanana, P. J. Kelly, and G. E. W. Bauer,
First-principles scattering matrices for spin transport,
\href{https://doi.org/10.1103/PhysRevB.73.064420}{Phys. Rev. B {\bf 73}, 064420 (2006).}

\bibitem{lu2021spin}
W.-T. Lu and Z. Yuan,
Spin accumulation and dissipation excited by an ultrafast laser pulse,
\href{https://doi.org/10.1103/PhysRevB.104.214404}{Phys. Rev. B {\bf 104}, 214404 (2021).}

\bibitem{lu2020interface}
W.-T. Lu, Y. Zhao, M. Battiato, Y. Wu, and Z. Yuan,
Interface reflectivity of a superdiffusive spin current in ultrafast demagnetization and terahertz emission,
\href{https://doi.org/10.1103/PhysRevB.101.014435}{Phys. Rev. B {\bf 101}, 014435 (2020).}

\bibitem{battiato2012theory}
M. Battiato, K. Carva, and P. M. Oppeneer,
Theory of laser-induced ultrafast superdiffusive spin transport in layered heterostructures,
\href{https://doi.org/10.1103/PhysRevB.86.024404}{Phys. Rev. B {\bf 86}, 024404 (2012).}

\bibitem{battiato2014treating}
M. Battiato, P. Maldonado, and P. M. Oppeneer,
Treating the effect of interface reflections on superdiffusive spin transport in multilayer samples (invited),
\href{https://doi.org/10.1063/1.4870589}{J. Appl. Phys. {\bf 115}, 172611 (2014).}

\bibitem{zhukov2005gw+}
V. P. Zhukov, E. V. Chulkov, and P. M. Echenique,
\emph{GW}+\emph{T} theory of excited electron lifetimes in metals,
\href{https://doi.org/10.1103/PhysRevB.72.155109}{Phys. Rev. B {\bf 72}, 155109 (2005).}

\bibitem{zhukov2006lifetimes}
V. P. Zhukov, E. V. Chulkov, and P. M. Echenique,
Lifetimes and inelastic mean free path of low-energy excited electrons in Fe, Ni, Pt, and Au: \emph{Ab initio} \emph{GW}+\emph{T} calculations,
\href{https://doi.org/10.1103/PhysRevB.73.125105}{Phys. Rev. B {\bf 73}, 125105 (2006).}

\bibitem{jiang2022ultrafast}
T. Jiang, X. Zhao, Z. Chen, Y. You, T. Lai, and J. Zhao,
Ultrafast enhancement and optical control of magnetization in ferromagnet/semiconductor layered structures via superdiffusive spin transports,
\href{https://doi.org/10.1016/j.mtphys.2022.100723}{Mater. Today Phys. {\bf 26}, 100723 (2022).}

\end{thebibliography}
\end{document}